\documentclass
[preprint,prl,twocolumn,amsfonts,amssymb,byrevtex,tightenlines,10pt,nobalancelastpage]{revtex4}%
\usepackage{amsfonts}
\usepackage{amsmath}
\usepackage{amssymb}
\usepackage{graphicx}%
\providecommand{\U}[1]{\protect\rule{.1in}{.1in}}

\begin{document}
\preprint{ }
\title{Evidence of Spatially Inhomogeous Pairing on the Insulating Side of a
Disorder-Tuned Superconductor-Insulator Transition}
\author{K. H. Sarwa B. Tan, Kevin A. Parendo, and A. M. Goldman}
\affiliation{School of Physics and Astronomy, University of Minnesota, Minneapolis, MN
55455, USA}
\pacs{PACS number}

\begin{abstract}
Measurements of transport properties of amorphous insulating In$_{x}$O$_{y}$
thin films have been interpreted as evidence of the presence of
superconducting islands on the insulating side of a disorder-tuned
superconductor-insulator transition. Although the films are not granular, the
behavior is similar to that observed in granular films. The results support
theoretical models in which the destruction of superconductivity by disorder
produces spatially inhomogenous pairing with a spectral gap.

\end{abstract}
\volumeyear{2007}
\volumenumber{number}
\issuenumber{number}
\eid{identifier}
\date[Date text]{date}
\received[Received text]{date}

\revised[Revised text]{date}

\accepted[Accepted text]{date}

\published[Published text]{date}

\startpage{1}
\endpage{4}
\maketitle

The interplay between localization and superconductivity can be investigated
through studies of disordered superconducting films \cite{GoldmanMarkovic},
originally treated by Anderson \cite{Anderson}, and Abrikosov and Gor'kov
\cite{AbrikosovGorkov}, who considered the low-disorder regime. Several
approaches have been proposed for strong disorder, including fermionic mean
field theories \cite{Fukuyama,MaLee,Finkelstein} and theories that focus on
the universal critical properties near the superconductor-insulator
transition. The latter consider the transition to belong to the dirty boson
universality class \cite{MPAFisher}. When quantum fluctuations are included in
fermionic theories for high levels of disorder a spatially inhomogeneous
pairing amplitude is found which retains a nonvanishing spectral gap
\cite{Trivedi}. For sufficiently disordered systems inhomogeneous pairing can
also be brought about by thermal fluctuations \cite{Bulaevskii}. A similar
inhomogeneous regime has also been considered under the rubric of electronic
microemulsions in the context of the metal-insulator transition of two
dimensional electron gases \cite{SpivakKivelson}. In this letter we provide
evidence of a spatially inhomogeneous order parameter on the insulating side
of a superconductor-insulator transition driven by structural and/or chemical disorder.

Studies of disorder and magnetic field tuned superconductor-insulator
transitions have usually been carried out on films that are either amorphous
or granular. In the former the disorder is on an atomic scale, and in the
latter, on a mesoscopic scale in which case the films consist of metallic
grains or clusters connected by tunneling, that are either embedded in an
insulating matrix, or on a bare substrate \cite{GoldmanMarkovic}. Amorphous
films can be produced when films of metal atoms such as Pb or Bi are grown at
liquid helium temperatures on substrates precoated with a wetting layer of
amorphous Ge or Sb \cite{Strongin}, or by careful vapor deposition of Mo$_{x}%
$Ge$_{y}$, In$_{x}$O$_{y}$, or TiN using a variety of techniques.

Granular films, are known to develop superconductivity in stages. \ If the
grains are small and weakly connected, the film is an insulator. For grains
larger than some characteristic size, and sufficiently close together,
\textquotedblleft local superconductivity\textquotedblright\ will develop
below some temperature. The opening of a spectral gap in the density of states
of the grains \cite{Orr1} results in a relatively sharp upturn in the
resistance below this temperature, which is usually close to the transition
temperature of the bulk material. \ For well enough coupled grains, there may
be a small drop in resistance at that temperature, followed by this upturn.
This is in contrast with the \textquotedblleft global
superconductivity\textquotedblright\ that occurs when a sufficient fraction of
the grains or clusters are strongly enough Josephson coupled to form a
percolating superconducting path across the film.

We have measured the temperature and magnetic field dependence of the
resistance, and nonlinear conductance-voltage characteristics of amorphous
In$_{x}$O$_{y}$ films prepared by electron-beam evaporation. These films are
not granular, but nevertheless exhibit local superconductivity at the lowest
temperatures. At low temperatures, the application of a magnetic field results
in a dramatic rise in resistance exhibiting a maximum that with decreasing
temperatures is found at decreasingly small fields. The conductance-voltage
characteristics in this high resistance regime are nonlinear in a manner
suggestive of single-particle tunneling between superconductors. We argue that
these observations are consistent with the presence of droplets, or islands of
superconductivity, characterized by a nonvanishing superconducting pair
amplitude and coupled by tunneling. Many of the droplets are Josephson
coupled, but their density is not high enough to produce a superconducting
path across the film.

The 22 nm\ thick films used in this study were deposited at a rate of 0.4 nm/s
by electron beam evaporation onto (001) SrTiO$_{3}$ epi-polished single
crystal substrates. Platinum electrodes, 10 nm\ in thickness, were deposited
prior to growth. The starting material was 99.999 \% pure In$_{2}$O$_{3}$.\ A
shadow mask defined a Hall bar geometry in which the effective area for
four-terminal resistance measurements was 500 x 500 $\mu$m$^{2}$. As-grown
films exhibited sheet resistances of about 4600 $\Omega$ at room temperature
and about 23 k$\Omega$ at 10 K. By annealing at relatively low temperatures
(55-70 $%
{{}^\circ}%
$C) in a high vacuum environment (10$^{-7}$ Torr), film resistances were
lowered, and depending upon the annealing time either insulating or
superconducting behavior at low temperatures could be induced \cite{Ovadyahu:
anneal}. Low-temperature rather than high-temperature annealing avoids changes
in morphology that would result in granular or microcrystalline films. As
reported by Gantmakher, \textit{et al}. \cite{Gantmakher}, at room temperature
the resistances of annealed films were found to be unstable. However, at low
temperatures (40-1400 mK) and in vacuum, they were stable.\ The films of the
present study had resistances of 2600 $\Omega$ at room temperature.\ 

Film structure was studied using atomic force microscopy (AFM), X-ray
diffraction (XRD) analysis, and high resolution scanning electron microscopy
(SEM). From the XRD there was no indication of the presence of crystalline In
or In$_{2}$O$_{3}$. The SEM did not detect any In inclusions, and could be
correlated with AFM studies which revealed for a 22 nm thick film, roughness
in the form of surface features with a height of 8.5 nm, and with bases about
18 nm in diameter. The conclusion from these characterization efforts is that
the films were homogeneous and amorphous, and did not contain isolated grains
or In inclusions.

Measurements were carried out in an Oxford Kelvinox-25 dilution refrigerator
housed in a screen room, with measuring leads filtered at room temperature
using $\pi$-section filters and RC filters. For measurements of resistance,
the applied current was set in the range of 10-100 pA, to avoid the
possibility of heating. Figure 1 shows a plot of $R$ $(T)$ for two films which
were studied in detail. For each, $dR/dT$ is negative at the lowest
temperatures.\ In the case of Film 1 there is a local minimum in $R(T)$ at
about 350 mK.\ Both films exhibit a sharp upturn in $R(T)$ between 200 and 300
mK, with the effects to be discussed below, occurring for Film 1 at higher
temperatures than for Film 2. These behaviors are suggestive of a regime of
local superconductivity \cite{Orr1}.%

\begin{figure}
[ptb]
\begin{center}
\includegraphics[
trim=0.000000in 0.000000in 0.028809in 0.000000in,
height=2.6783in,
width=2.6498in
]%
{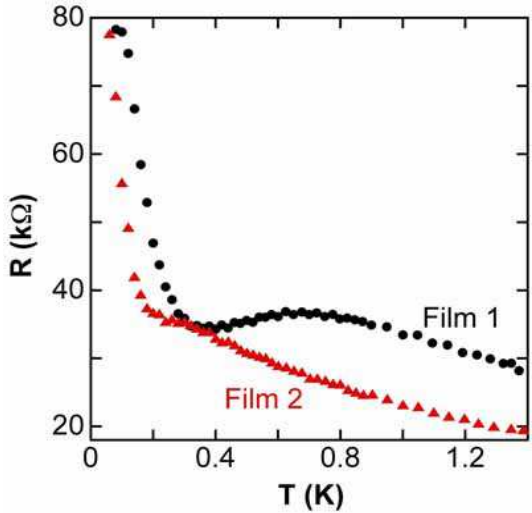}%
\caption{Resistance vs. temperature for Films 1 and 2.}%
\end{center}
\end{figure}

The sheet resistances of Films 1 and 2 were both approximately 78 k$\Omega$ at
40 mK. In a perpendicular magnetic field of only 0.2 T, their sheet
resistances increased by up to a factor of 40 at 40 mK. The maximum in $R(B)$
as shown in Fig. 2(a) for Film 1 is followed, at the lowest temperatures, by a
relatively slow decrease in resistance with increasing field. The resistance
maximum moves to higher fields, with increasing temperature. The behavior of
Film 2 resembled the higher temperature data for Film 1, presumably because
Film 2 exhibited weaker traces of superconductivity as evidenced by the
absence of a local minimum in $R(T)$ in the zero-field. This variation in
properties from film to film is expected, as small changes in chemistry and/or
morphology can have a large effect on disordered film properties. The
temperature dependencies of the fields, $B_{peak}$ and resistances $R_{peak}$
are presented in Fig. 2(b). A qualitatively similar, but weaker enhancement of
resistance was previously reported for insulating In$_{x}$O$_{y}$ films by
Gantmakher and coworkers \cite{Gantmakher}. A larger enhancement was reported
for ultrathin insulating Be thin films \cite{Wu: Be film}. However neither of
these works demonstrated the systematic effects shown in Fig. 2(b).%

\begin{figure}
[ptb]
\begin{center}
\includegraphics[
trim=0.000000in 0.000000in -0.115168in 0.000000in,
height=4.5455in,
width=3.2603in
]%
{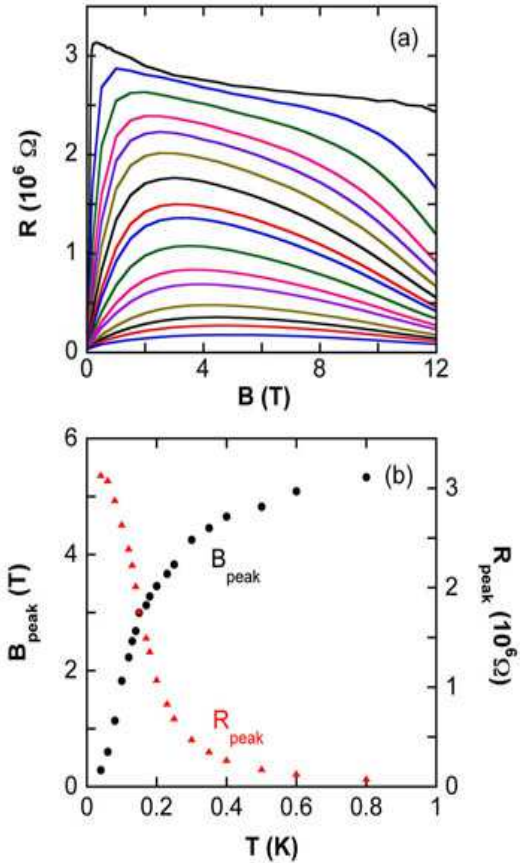}%
\caption{(a) Resistance vs. magnetic field for Film 1. The temperatures are 40
(top), 80, 100, 120, 130, 140, 150, 170, 180, 200, 230, 250, 300, 350, 400,
and 500 mK (bottom). (b)The fields (left axis) and the resistances (right
axis) of the peaks in $R(B)$ are plotted as a function of temperature. The
flattening of $R_{peak\text{ }}$at the lowest temperatures may be the result
of a failure to cool the electrons. }%
\end{center}
\end{figure}

To probe the nature of the high resistance state, differential
conductance-voltage characteristics were also studied \cite{Shahar:
SC,Christiansen,Baturina1,Wu: Be film,Wu: Al film, Barber}. These are shown in
Fig. 3 for Film 2 which was studied in detail. Film 1 exhibited qualitatively
similar features.%

\begin{figure}
[ptb]
\begin{center}
\includegraphics[
trim=0.000000in 0.000000in -0.277542in 0.000000in,
height=2.4811in,
width=3.0831in
]%
{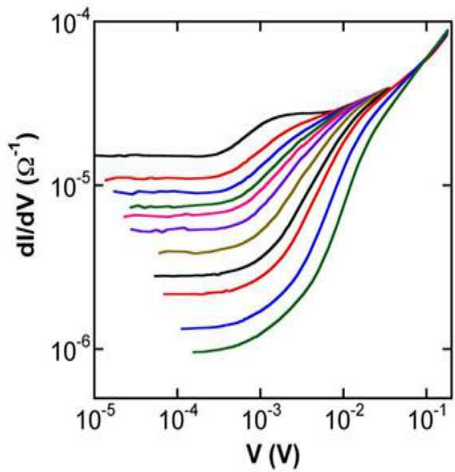}%
\caption{Differential conductance vs. voltage of Film 2 at 100 mK for 0 (top),
0.01, 0.02, 0.03, 0.04, 0.06, 0.1, 0.175, 0.25, 0.5, and 1 T (bottom).}%
\end{center}
\end{figure}
All of the nonlinear conductance-voltage characteristics are reminiscent of
the single-particle tunneling characteristics of
superconductor-insulator-superconductor (SIS) tunneling junctions. The effects
of electron heating are found at voltages well above the observed conductance
thresholds \cite{Gershenson}. The fact that the low voltage nonlinearities
vanish at temperatures above approximately 200 mK, suggests that they are
associated with the presence of a nonvanishing pairing amplitude occurring in
the disconnected superconducting regions.

We can model these thin disordered films exhibiting spatially inhomogeneous
pairing as random networks of tunneling junctions of various (random) levels
of conductivity, connecting superconducting clusters imbedded in an insulating
matrix. Some of these junctions are superconducting because they are Josephson
coupled. As a consequence there are \textquotedblleft
superclusters,\textquotedblright\ which are aggregates of Josephson coupled
smaller clusters that may cover a macroscopic fraction of the film area.
Charge may flow through \textquotedblleft superclusters\textquotedblright%
\ with zero electrical resistance. However, as long as these do not span the
film the resistance will be determined by single-particle tunneling. The sheet
resistance of the resultant network can be inferred using the following simple
argument \cite{Orr2}. Disconnect all of the junctions in the network whose
conductance involves single particle tunneling, and then reconnect them one by
one in ascending order of resistance. A stage will be reached at which the
next junction completes an infinite cluster connecting the ends of the
network. Let the normal state resistance of this last junction be $R_{n}$. The
actual value will depend upon the nature of the distribution of
single-particle tunneling resistances in the film. The measured normal-state
sheet resistance of the entire network will then be $R_{n}$, as this junction
is the bottleneck. Junctions with $R>R_{n}$ are irrelevant since they are
always shunted by junctions with resistances of order $R_{n}$. Junctions with
$R<R_{n}$ only form finite clusters over macroscopic distances. They don't
affect the conductivity because current must still pass through junctions with
resistance of order $R_{n}$ to get from one \textquotedblleft
supercluster\textquotedblright\ to the next. The action of a magnetic field is
to quench the Josephson coupling within the \textquotedblleft
superclusters.\textquotedblright\ When this happens, the resistance at each
temperature will be governed by the new, higher, value of the bottleneck
resistance as there will no longer be any Josephson-coupled \textquotedblleft
superclusters,\textquotedblright\ and the distribution of junction resistances
will shift towards higher values of resistance.

The fact that the magnetic field that induces higher resistance decreases with
decreasing temperature is a counter-intuitive result, implying a divergent
magnetic length scale in the zero temperature limit, possibly of the form
$\left[  \phi_{0}/B\right]  ^{1/2}$where $\phi_{0}$ is the flux quantum. This
result suggests enhanced quantum fluctuations in the zero temperature limit. A
heuristic argument can be made to demonstrate that this is plausible by
treating the inhomogeneous pairing state of the film as a collection of
superconducting grains or islands coupled by tunneling junctions. Without the
inclusion of quantum fluctuations the argument may not capture all of the
features of the data. It is first useful to consider the magnetic field
dependence of the in-plane Josephson coupling between two planar thin film
square islands with an area $L^{2}$. This is a geometry resembling the grain
boundary geometry of high temperature superconductor junctions. A magnetic
field applied perpendicular to the plane will completely penetrate both
electrodes of such a junction. As a consequence the minima of the diffraction
pattern will be governed by the field corresponding to a single flux quantum
over the full area of the structure or $B\left[  L(2L+d)\right]  =\phi_{0}$
where d is the width of the barrier or gap \cite{Ngai}. Since $L>>d$, the
field at the first minimum of the diffraction pattern would be found at a
value proportional to $\phi_{0}/L^{-2}$ . For a \textquotedblleft
supercluster\textquotedblright\ consisting of a square array of square islands
that are Josephson coupled, with some degree of randomness in the coupling,
one would expect coherence to vanish at the first minimum. For a random array
and with clusters that are not square, one might expect a similar dependence
on $L^{-2}$. If the characteristic size of the islands increased with
decreasing temperature, which is a plausible assumption, the field quenching
the Josephson coupling, would be expected to decrease as is observed. For the
films studied, at the lowest temperatures, the peak in the resistance occurs
in a field of 0.2 T, which would correspond to a length of approximately 100 nm.

The fall off of the resistance at fields above that producing the maximum
slows with decreasing temperature, consistent with the strengthening of the
pairing amplitude with decreasing temperature. The fact that this remnant of
superconductivity persists to fields up to 12 T, far above the bulk critical
field of In$_{x}$O$_{y}$, implies that the superconducting islands are much
smaller than the penetration depth. It should be possible to develop a
detailed percolation model of this effect, similar to that developed for
granular superconductors \cite{Pury}, which includes the quenching of the
Josephson coupling by magnetic field and quantum fluctuations.

Although the resistance of the films of the present work increases with
decreasing temperatures in zero magnetic field there is no guarantee that at
some temperature lower than the minimum value accessed in these measurements,
the resistance will not fall to zero. This could result from the percolation
of Josephson coupling across the film as the size of the clusters increases.
In that event the inhomogeneous pairing implied by the data would be more
likely governed by a theory including both quantum \cite{Trivedi} and thermal
\cite{Bulaevskii} fluctuations.

The large peaks in the magnetoresistance found at fields above the magnetic
field-induced superconductor-insulator transition of superconducting amorphous
In$_{x}$O$_{y}$ thin films may result from a similar inhomogeneity of the
pairing amplitude, in that case induced by magnetic field rather than
disorder. Such peaks were first reported by Hebard and Paalanen \cite{Hebard:
SIT} who suggested that the state induced when superconductivity was quenched
was a Bose insulator, characterized by localized Cooper pairs. They proposed
that the peak was the signature of a crossover to a Fermi insulating state of
localized electrons. This resistance peak has been the subject of more recent
studies of In$_{x}$O$_{y}$ films \cite{Gantmakher,Shahar: SC,Steiner}, of
microcrystalline TiN films where the high field limit appears to be a
\textquotedblleft quantum metal\textquotedblright\ \cite{Baturina1},\ and of
high-$T_{c}$ superconductors \cite{Kapitulnik: high Tc}. The fact that
inhomogeneous pairing can be induced in disordered superconductors by magnetic
fields has been recently established using a Hubbard Model \cite{Dubi}.

The notion that disorder implies inhomogeneity of superconducting order on
some length scale was first discussed by Kowal and Ovadyahu
\cite{KowalOvadyahu}, and as was mentioned earlier emerges naturally in a
fermionic model of the superconductor-insulator transition that exhibits a
disorder-tuned inhomogeneity of the pairing amplitude \cite{Trivedi}. The
films of Kowal and Ovadyahu differ from those of the present work in that they
are presumably more disordered, and thus further into the insulating regime.
Their magnetoresistance is always negative as there is no Josephson coupling
between islands and the main effect of magnetic field is to weaken the
inhomogeneous pairing amplitude, leading to negative magnetoresistance. \ 

This work was supported by the National Science Foundation under grant no.
NSF/DMR-0455121. The authors would like to thank Zvi Ovadyahu for providing
samples and for critical comments, and Leonid Glazman and Alex Kamenev for
useful discussions.

\end{document}